\documentclass[a4paper,onecolumn,12pt]{IEEEtran}
\usepackage{cite,graphicx,amsmath,amssymb}
\usepackage{subfigure}
\usepackage{citesort}
\usepackage{fancyhdr}
\usepackage[justification=centering]{caption}
\usepackage{balance}
\usepackage[utf8x]{inputenc}
\usepackage{epstopdf}
\usepackage{epsfig}	
\usepackage{xcolor}
\newtheorem{theorem}{Theorem}

\newtheorem{lemma}{Lemma}

\newtheorem{corollary}{Corollary}

\begin{document}

\title{Non-orthogonal Multiple Access in Large-Scale Underlay Cognitive Radio Networks}
\author{Yuanwei\ Liu, Zhiguo\ Ding, Maged\ Elkashlan, and Jinhong\ Yuan
\thanks{Y. Liu and M. Elkashlan are with Queen Mary University of London, London,
UK (email: \{yuanwei.liu, maged.elkashlan\}@qmul.ac.uk).}
\thanks{ Z. Ding is with Lancaster University, Lancaster, UK (e-mail: z.ding@lancaster.ac.uk).}
\thanks{ J. Yuan is with the University of New South Wales, Sydney, Australia (e-mail: j.yuan@unsw.edu.au).}
}
\maketitle
\vspace{-2cm}
\begin{abstract}
In this paper, non-orthogonal multiple access (NOMA) is applied to large-scale underlay cognitive radio (CR) networks with randomly deployed users. In order to characterize the performance of the considered network, new closed-form expressions of the outage probability are derived using stochastic-geometry. More importantly, by carrying out  the diversity analysis, new insights are obtained under the two  scenarios with different power constraints: 1) fixed transmit power of the primary transmitters (PTs), and 2) transmit power of the PTs being proportional to that of the secondary base station. For the first scenario, a diversity order of $m$ is experienced at the $m$-th ordered NOMA user. For the second scenario, there is an asymptotic error floor for the outage probability. Simulation results are provided to verify  the accuracy of the derived results. A pivotal conclusion is reached that by carefully designing target data rates and power allocation coefficients of users, NOMA can outperform conventional orthogonal multiple access in underlay CR networks.
\end{abstract}

\vspace{-1cm}
\section{Introduction}
Spectrum efficiency is of significant importance and becomes one of the main design targets for future fifth generation networks. Non-orthogonal multiple access (NOMA) has received considerable attention because of  its potential to achieve superior spectral efficiency \cite{saito2013system}. Particularly, different from  conventional multiple access (MA) techniques, NOMA uses the power domain to serve multiple users at different power levels in order to use spectrum more efficiently. A downlink NOMA and an uplink NOMA are considered in~\cite{ding2014performance} and \cite{al2014uplink}, respectively. The application of multiple-input multiple-output (MIMO) techniques to NOMA has been considered in \cite{ding2015mimo} by using zero-forcing detection matrices. The authors in \cite{SunqiWCL} investigated an ergodic capacity maximization problem for MIMO NOMA systems.

Another approach to improve spectrum efficiency is the paradigm of underlay cognitive radio (CR) networks, which was proposed in \cite{goldsmith2009breaking} and  has rekindled increasing interest in using spectrum more efficiently. The key idea of underlay CR networks is that each secondary user (SU) is allowed to access the spectrum of the primary users (PUs) as long as the SU meets a certain interference threshold   in the primary network (PN). In \cite{7037366}, an underlay CR network taking into account the spatial distribution of the SU relays and PUs was considered and its performance was evaluated by using stochastic geometry tools.  In \cite{ding2014pairing}, a new  CR inspired NOMA scheme has been proposed and the impact of user pairing has been  examined, by focusing on a simple scenario with only one primary transmitter (PT).

By introducing the aforementioned two concepts, it is natural to consider the application of NOMA in underlay CR networks using additional power control at the secondary base station (BS) to improve the spectral efficiency. Stochastic geometry is used to model a large-scale CR network with a large number of randomly deployed PTs and primary receivers (PRs). We consider a practical system design as follows: 1) All the SUs, PTs, and PRs are randomly deployed based on the considered stochastic geometry model; 2) Each SU suffers interference from other NOMA SUs as well as the PTs; and 3) The secondary BS must satisfy a predefined power constraint threshold to avoid interference at the PRs. New closed-form expressions of the outage probability of the NOMA users are derived to evaluate the performance of the considered CR NOMA network. Moreover, considering two different power constraints at the PTs, diversity order\footnote{Diversity order is defined as the slope for the outage provability curve decreasing with the signal-to-noise-ratio (SNR). It measures the number of independent fading paths over which the data is received. In NOMA networks, since users' channels are ordered and SIC is applied at each receiver, it is of importance to investigate the diversity order.} analysis is carried out with providing important insights: 1) When the transmit power of the PTs is fixed, the $m$-th user among all ordered NOMA user experiences a diversity order of $m$; and 2) When the the transmit power of the PTs is proportional to that of the secondary BS, an asymptotic error floor exists for the outage probability.
\section{Network Model}
We consider a large-scale underlay spectrum sharing scenario consisting of the PN and the secondary network (SN). In the SN, we consider that a   secondary BS is located at the origin of a disc, denoted by $\mathcal{D}$ with radius ${R_{{D}}}$ as its coverage. The $M$ randomly deployed secondary users are uniformly distributed within the disc which is the user zone for NOMA. The secondary BS communicates with all SUs within the disc by applying the NOMA transmission protocol. It is worthy pointing out that the power of the secondary transmitter is constrained in order to limit the interference at the PRs. In the PN, we consider a random number of PTs and PRs distributed in an infinite two dimensional plane. The spatial topology of all the PTs and PRs are modeled using homogeneous poisson point processes (PPPs), denoted by ${\Phi_b}$ and ${\Phi _\ell}$ with density ${\lambda_b}$ and ${\lambda_\ell}$, respectively. All channels are assumed to be quasi-static Rayleigh fading   where the channel coefficients are constant for each transmission block but vary independently between different blocks.

According to underlay CR,  the transmit power ${P_t}$ at the secondary BS is constrained as follows:
\begin{align}\label{Transmit power}
{P_t} = \min \left\{ {\frac{{{I_p}}}{{\mathop {\max }\limits_{\ell  \in {\Phi _\ell }} {{\left| {{g_\ell }} \right|}^2}}},P_{s}} \right\},
\end{align}
where $I_p$ is the maximum permissible interference power at the PRs, ${P_s}$ is maximum transmission power at the secondary BS, ${\left| {{g_\ell }} \right|^2} = {\left| {{{\hat g}_\ell }} \right|^2}L\left( {{d_\ell }} \right)$ is the overall channel gain from the secondary BS to PRs $\ell$. Here,  ${{{\hat g}_\ell }}$ is small-scale fading with ${{\hat g}_\ell } \sim \mathcal{CN}( {0,1})$, $L\left( {{d_\ell }} \right) = \frac{1}{1+{d_\ell ^\alpha }}$ is large-scale path loss, ${d_\ell}$ is the distance between the secondary BS and the PRs, and $\alpha$ is the path loss exponent. A bounded path loss model is used to ensure the path loss is always larger than one even for small distances~\cite{ding2014performance,stochastic}.

According to NOMA, the BS sends a combination of messages to all NOMA users, and the observation at the $m$-th secondary user is given by
\begin{align}\label{reveiving signal SU}
{y_m} = {h_m}\sum\limits_{n = 1}^M {\sqrt {{a_n}{P_t}} {x_n}}  + {n_m},
\end{align}
where  $n_m$ is the additive white Gaussian noise (AWGN) at the $m$-th user with variance $\sigma^2$, ${{a_n}}$ is the power allocation coefficient for the $n$-th SU with $\sum\nolimits_{n = 1}^M {{a_n}}  = 1$, ${{x_n}}$ is the information for the $n$-th user, and $h_m$ is the channel coefficient between the $m$-th user and the secondary BS.

For the SUs, they also observe the interferences of the randomly deployed PTs in the PN.
Usually, when the PTs are close to the secondary NOMA users, they will cause significant interference. To overcome this issue, we introduce  an interference guard zone $D_0$ to each secondary NOMA user with radius of $d_0$, which means that there is no interference from PTs allowed inside this zone \cite{venkataraman2006shot}. We assume $d_0 \ge 1$ in this paper. The interference links from the PTs to the SUs are dominated by the path loss and is given by
\begin{align}\label{Interference}
{I_B} = \sum\limits_{b \in {\Phi _b}} {L\left( {{d_b}} \right)}  ,
\end{align}
where $L(d_b)=1/(1+d_b^{\alpha})$ is the large-scale path loss and ${d_b}$ is the distance from the PTs to the SUs.

Without loss of generality, all the channels of SUs are assumed to follow the order as ${\left| {{h_1}} \right|^2} \le {\left| {{h_2}} \right|^2} \le  \cdot  \cdot  \cdot  \le {\left| {{h_M}} \right|^2}$. The power allocation coefficients are assumed to follow the order as ${a_1} \ge {a_2} \ge  \cdot  \cdot  \cdot  \ge {a_M}$. According to the NOMA principle, successive interference cancelation (SIC) is carried out at the receivers \cite{cover2006elements}. It is assumed that $1 \le j \le m < i$. In this case, the $m$-th user can decode the message of the $j$-th user and treats the message for the $i$-th user as interference. Specifically, the $m$-th user first decodes the messages of all the $(m-1)$ users, and then successively subtracts these messages to obtain its own information. Therefore, the received signal-to-interference-plus-noise ratio (SINR) for the $m$-th user to decode the information of the $j$-th user is given by
\begin{align}\label{SINR m i}
&{\gamma _{m,j}}
= \frac{{{{\left| {{h_m}} \right|}^2}{\gamma _t}{a_j}}}{{{{\left| {{h_m}} \right|}^2}{\gamma _t}\sum\limits_{i = j + 1}^M {{a_i}}  + {\rho _b}{I_B}  + 1}},
\end{align}
where ${\gamma _t} = \min \left\{ {\frac{{{\rho _p}}}{{\mathop {\max }\limits_{\ell  \in {\Phi _\ell }} {{\left| {{g_\ell }} \right|}^2}}},{\rho _s}} \right\}$, ${\rho _p} = \frac{{{I_p}}}{{{\sigma ^2}}},{\rho _s} = \frac{{P_{s}}}{{{\sigma ^2}}},{\rho _b} = \frac{{{P_B}}}{{{\sigma ^2}}}$, and ${P_B}$ is the transmit power of the PTs, ${\left| {{h_m}} \right|^2}$ is the overall ordered channel gain from the secondary BS to the $m$-th SU. For the case $m=j$, it indicates the $m$-th user decodes the message of itself. Note that the SINR for the $M$-th SU is ${\gamma _{M,M}} = \frac{{{{\left| {{h_M}} \right|}^2}{\gamma _t}{a_M}}}{{{\rho _b}{I_B} + 1}}$.

\section{Outage Probability}
In this section, we provide exact analysis of the considered networks in terms of outage probability. In NOMA, an outage occurs if the $m$-th user can not detect any of the $j$-th user's message, where $j \le m$ due to the SIC. Denote ${X_m} = \frac{{{{\left| {{h_m}} \right|}^2}{\gamma _t}}}{{{\rho _b}{I_B} + 1}}$. Based on \eqref{SINR m i}, the cumulative distribution function (CDF) of $X_m$ is given by
\begin{align}\label{CDF_X_m}
{F_{{X_m}}}\left( {{\varepsilon}} \right) = \Pr \left\{ {\frac{{{{\left| {{h_m}} \right|}^2}{\gamma _t}}}{{{\rho _b}{I_B} + 1}} < {\varepsilon}} \right\}.
\end{align}
We denote ${\varepsilon _j}={\tau _j}/\left( {{a_j} - {\tau _j}\sum\nolimits_{i = j + 1}^M {{a_i}} } \right)$ for $j<M$, ${\tau _j} = {2^{{R_j}}} - 1$, $R_j$ is the target data rate for the $j$-th user,  ${\varepsilon _M} = {\tau _M}/{a_M}$, and $\varepsilon _m^{\max } = \max \left\{ {{\varepsilon _1},{\varepsilon _2},...,{\varepsilon _m}} \right\}$. The outage probability at the $m$-th user can be expressed as follows:
\begin{align}\label{Pout exact 1}
{P_m} = \Pr \left\{ {{X_m} < \varepsilon _m^{\max }} \right\} = {F_{{X_m}}}\left( {\varepsilon _m^{\max }} \right),
\end{align}
where the condition ${a_j} - {\tau _j}\sum\nolimits_{i = j + 1}^M {{a_i}}  > 0$ should be satisfied due to applying NOMA, otherwise the outage probability will always be one \cite{ding2014performance}.
%

We need calculate the CDF of $X_m$ conditioned on ${I_B}$ and ${\gamma _t}$. Rewrite \eqref{CDF_X_m} as follows:
\begin{align}\label{CDF_X_m_1}
{F_{\left. {{X_m}} \right|{I_B},{\gamma _t}}}\left( \varepsilon  \right) = {F_{{{\left| {{h_m}} \right|}^2}}}\left( {\frac{{\left( {{\rho _b}{I_B} + 1} \right)\varepsilon }}{{{\gamma _t}}}} \right),
\end{align}
where ${F_{{{\left| {{h_m}} \right|}^2}}}$ is the CDF of $h_m$. Based on order statistics \cite{order} and applying binomial expansion, the CDF of the ordered channels has a relationship with the unordered channels as follows:
\begin{align}\label{order stacastics}
{F_{{{\left| {{h_m}} \right|}^2}}}\left( y \right) = {\psi _m}\sum\limits_{p = 0}^{M - m} {
 M - m \choose
 p} \frac{{{{\left( { - 1} \right)}^p}}}{{m + p}}{\left( {{F_{{{\left| {{{\tilde h}}} \right|}^2}}}\left( y \right)} \right)^{m + p}},
\end{align}
where $y = \frac{{\left( {{\rho _b}{I_B} + 1} \right){\varepsilon}}}{{{\gamma _t}}}$
, ${\psi _m} = \frac{{M!}}{{\left( {M - m} \right)!\left( {m - 1} \right)!}}$, and ${{\left| {{{\tilde h}}} \right|}^2}={\left| {{{\hat h}}} \right|^2}L\left( {{d}} \right)$ is the unordered channel gain of an arbitrary SU. Here,  ${{{\hat h}}}$ is the small-scale fading coefficient with ${{{{\hat h}}}} \sim \mathcal{CN}( {0,1})$, $L\left( {{d}} \right)= \frac{1}{1+{d ^\alpha }}$ is the large-scale path loss, and $d$ is a random variable representing the distance from the secondary BS to an arbitrary SU.

Then using the assumption of homogenous PPP and applying the polar coordinates, we express ${F_{{{\left| {{{\tilde h}}} \right|}^2}}}\left( y \right) $ as follows:
\begin{align}\label{CDF_X_m_wave}
{F_{{{\left| {{{\tilde h}}} \right|}^2}}}\left( y \right) = \frac{2}{{R_D^2}}\int_0^{{R_D}} {\left( {1 - {e^{ - \left( {1 + {r^\alpha }} \right)y}}} \right)rdr}.
\end{align}
Note that it is challenging to obtain an insightful expression for the unordered CDF. As such, we apply the Gaussian-Chebyshev quadrature \cite{Hildebrand1987introduction} to find an approximation for \eqref{CDF_X_m_wave} as
\begin{align}\label{CDF_X_m_wave_GC}
{F_{{{\left| {{{\tilde h}}} \right|}^2}}}\left( y \right) \approx \sum\limits_{n = 0}^N {{b_n}{e^{ - {c_n}y}}},
\end{align}
where $N$ is a complexity-accuracy tradeoff parameter, ${b_n} =  - {\omega _N}\sqrt {1 - \phi _n^2} \left( {{\phi _n} + 1} \right)$, ${b_0} =  - \sum\limits_{n = 1}^N {{b_n}}$, ${c_n} = 1+ {\left( {\frac{{{R_D}}}{2}\left( {{\phi _n} + 1} \right)} \right)^\alpha }$, ${\omega _N} = \frac{\pi }{N}$, and ${\phi _n} = \cos \left( {\frac{{2n - 1}}{{2N}}\pi } \right)$.

Substituting \eqref{CDF_X_m_wave_GC} into \eqref{order stacastics} and applying the multinomial theorem, we obtain
\begin{align}\label{CDF_X_m_2}
{F_{{{\left| {{h_m}} \right|}^2}}}\left( y \right)= {\psi _m}\sum\limits_{p = 0}^{M - m} {
 M - m \choose
 p  } \frac{{{{\left( { - 1} \right)}^p}}}{{m + p}}\sum\limits_{{q_0} +  \cdots  + {q_N} = m + p} {{
 m + p \choose
 {q_0} +  \cdots  + {q_N}}\left( {\prod\limits_{n = 0}^N {b_n^{{q_n}}} } \right)}{e^{ - \sum\limits_{n = 0}^N {{q_n}{c_n}y} }}.
\end{align}
where $
{ m + p \choose
 {q_0} +  \cdots  + {q_N} } = \frac{{\left( {m + p} \right)!}}{{{q_0}! \cdots {q_N}!}}$. Based on \eqref{CDF_X_m_2}, the CDF of $X_m$ can be expressed as follows:
\begin{align}\label{CDF_X_m_wave_GC_1}
{F_{{X_m}}}\left( {{\varepsilon _j}} \right) =& \int_0^\infty  {\int_0^\infty  {{F_{{{\left| {{h_m}} \right|}^2}}}\left( {\frac{{\left( {{\rho _b}x + 1} \right){\varepsilon _j}}}{z}} \right){f_{{I_B}}}\left( x \right){f_{{\gamma _t}}}\left( z \right)dxdz} }  \nonumber\\
 =& {\psi _m}\sum\limits_{p = 0}^{M - m} {
 M - m \choose
 p } \frac{{{{\left( { - 1} \right)}^p}}}{{m + p}} \sum\limits_{{q_0} +  \cdots  + {q_N} = m + p} {{
 m + p \choose
 {q_0} +  \cdots  + {q_N}
}\left( {\prod\limits_{n = 0}^N {b_n^{{q_n}}} } \right)}\nonumber\\
 &\times \underbrace {\int_0^\infty  {{e^{ - \frac{{{\varepsilon _j}}}{z}\sum\limits_{n = 0}^N {{q_n}{c_n}} }}\underbrace {\int_0^\infty  {{e^{ - \frac{{x{\rho _b}{\varepsilon _j}}}{z}\sum\limits_{n = 0}^N {{q_n}{c_n}} }}{f_{{I_B}}}\left( x \right)} dx}_{{Q_2}}{f_{{\gamma _t}}}\left( z \right)} dz}_{{Q_1}},
\end{align}
where ${f_{{{\gamma _t}}}}$ is the PDF of ${\gamma _t}$. We express $Q_2$ in \eqref{CDF_X_m_wave_GC_1} as follows:
\begin{align}\label{Q2_1}
{Q_2} = \int_0^\infty  {{e^{ - x\frac{{{\rho _b}{\varepsilon _j}}}{z}\sum\limits_{n = 0}^N {{q_n}{c_n}} }}{f_{{I_B}}}\left( x \right)} dx= {E_{{\Phi _b}}}\left\{ {{e^{ - \frac{{x{\rho _b}{\varepsilon _j}}}{z}\sum\limits_{n = 0}^N {{q_n}{c_n}} }}} \right\} = {L_{{I_B}}}\left( {\frac{{x{\rho _b}{\varepsilon _j}}}{z}\sum\limits_{n = 0}^N {{q_n}{c_n}} } \right).
\end{align}

In this case, the Laplace transformation of the interferences from the PT can be expressed as\cite{venkataraman2006shot}
\begin{align} \label{Lapace_I_B guard zone}
{L_{I_B}}\left( s \right)& =\exp \left( { - {\lambda _b}\pi \left[ {\left( {{e^{ - sd_0^{ - \alpha }}} - 1} \right)d_0^2 + {s^\delta }\gamma \left( {1 - \delta ,sd_0^{ - \alpha }} \right)} \right]} \right)\nonumber\\
 &= \exp \left( { - {\lambda _b}\pi \left[ {\left( {{e^{ - sd_0^{ - \alpha }}} - 1} \right)d_0^2 + {s^\delta }\underbrace {\int_0^{sd_0^{ - \alpha }} {{t^{ - \delta }}{e^{ - t}}dt} }_\Theta } \right]} \right),
\end{align}
where $\delta  = \frac{2}{\alpha }$ and $\gamma \left( {\cdot} \right)$ is the lower incomplete Gamma function.

To obtain an insightful expression, we use Gaussian-Chebyshev quadrature to approximate the lower incomplete Gamma function in \eqref{Lapace_I_B guard zone}, $\Theta$ can be expressed as follows:
\begin{align} \label{Theta GC}
\Theta  \approx {s^{1 - \delta }}\sum\limits_{l = 1}^L {{\beta _l}{e^{ - {t_l}sd_0^{ - \alpha }}}},
\end{align}
where $L$ is a complexity-accuracy tradeoff parameter, ${\beta _l} = \frac{1}{2}d_0^{2 - \alpha }{\omega _L}\sqrt {1 - \theta _l^2} {t_l}^{ - \delta }$
, ${t_l} = \frac{1}{2}\left( {{\theta _l} + 1} \right)$, ${\omega _L} = \frac{\pi }{L}$, and ${\theta _l} = \cos \left( {\frac{{2l - 1}}{{2L}}\pi } \right)$. Substituting \eqref{Theta GC} into \eqref{Lapace_I_B guard zone}, we approximate the Laplace transformation as follows:
\begin{align} \label{Lapace_I_B guard zone GC}
{L_{I_B}}\left( s \right) \approx& {e^{ - {\lambda _b}\pi \left( {\left( {{e^{ - sd_0^{ - \alpha }}} - 1} \right)d_0^2 + {s}\sum\limits_{l = 1}^L {{\beta _l}{e^{ - {t_l}sd_0^{ - \alpha }}}} } \right)}}.
\end{align}

Substituting \eqref{Lapace_I_B guard zone GC} into \eqref{CDF_X_m_wave_GC_1}, $Q_2$ is given by
\begin{align} \label{CCDF GZ 1}
{Q_2} = {e^{ - {\lambda _b}\pi \left( {\left( {{e^{ - \frac{{{\rho _b}{\varepsilon _j}d_0^{ - \alpha }}}{z}\sum\limits_{n = 0}^N {{q_n}{c_n}} }} - 1} \right)d_0^2 + \frac{{{\rho _b}{\varepsilon _j}}}{z}\sum\limits_{n = 0}^N {{q_n}{c_n}} \sum\limits_{l = 1}^L {{\beta _l}{e^{ - \frac{{{t_l}{\rho _b}{\varepsilon _j}}}{{zd_0^\alpha }}\sum\limits_{n = 0}^N {{q_n}{c_n}} }}} } \right)}}.
\end{align}

The following theorem provides the PDF of ${\gamma _t}$.
\begin{theorem}\label{theorem:A}
Consider the use of the composite channel model with Rayleigh fading and path loss, the PDF of the effective power of the secondary BS  is given by
\begin{align}\label{PDF Pt}
{f_{{\gamma _t}} }\left( x \right) ={e^{ - {a_\ell }\rho _s^\delta {e^{ - \frac{{{\rho _p}}}{x}}}}}Dirac\left( {x - {\rho _s}} \right)+ \left( {\frac{{{\rho _p}}}{x} + \delta } \right){a_\ell }{x^{\delta  - 1}}{e^{ - {a_\ell }{x^\delta }{e^{ - \frac{{{\rho _p}}}{x}}} - \frac{{{\rho _p}}}{x}}}U\left( {{\rho _s} - x} \right),
\end{align}
where ${a_\ell } = \frac{{\delta \pi {\lambda _\ell }\Gamma \left( \delta  \right)}}{{\rho _p^\delta }}$, $U\left( \cdot \right)$ is the unit step function, and $Dirac\left( \cdot \right)$ is the impulse function.
\begin{proof}
See Appendix~A.
\end{proof}
\end{theorem}

Substituting \eqref{PDF Pt} and \eqref{CCDF GZ 1} into \eqref{CDF_X_m_wave_GC_1}, we express $Q_1$ as follows:
\begin{align}\label{Q_1_temp}
{Q_1} =& {e^{ - {a_\ell }\rho _s^\delta {e^{ - \frac{{{\rho _p}}}{{{\rho _s}}}}} - \frac{{{\varepsilon _j}\sum\limits_{n = 0}^N {{q_n}{c_n}} }}{{{\rho _s}}} - {\lambda _b}\pi \left( {\left( {{e^{ - \frac{{{\rho _b}{\varepsilon _j}}}{{{\rho _s}d_0^\alpha }}\sum\limits_{n = 0}^N {{q_n}{c_n}} }} - 1} \right)d_0^2 + \frac{{{\rho _b}{\varepsilon _j}}}{{{\rho _s}}}\sum\limits_{n = 0}^N {{q_n}{c_n}} \sum\limits_{l = 1}^L {{\beta _l}{e^{ - \frac{{{t_l}{\rho _b}{\varepsilon _j}}}{{{\rho _s}d_0^\alpha }}\sum\limits_{n = 0}^N {{q_n}{c_n}} }}} } \right)}}   \nonumber\\
&+ \underbrace {\int_0^{{\rho _s}} {{a_\ell }\left( {\frac{{{\rho _p}}}{z} + \delta } \right){z^{\delta  - 1}}{e^{ - {a_\ell }{z^\delta }{e^{ - \frac{{{\rho _p}}}{z}}} - \frac{{{\rho _p} + {\varepsilon _j}\sum\limits_{n = 0}^N {{q_n}{c_n}} }}{z}}}{Q_2}} dz}_\Psi .
\end{align}

We notice that it is very challenging to solve the integral $\Psi$ in \eqref{Q_1_temp}, therefore, we apply the Gaussian-Chebyshev quadrature to approximate the integral as follows:
\begin{align}\label{GC Phi}
\Psi  \approx \sum\limits_{k = 1}^K {{\eta _k}} {e^{ - \frac{{{\rho _p} + {\varepsilon _j}\sum\limits_{n = 0}^N {{q_n}{c_n}} }}{{{\rho _s}{s_k}}} - {\lambda _b}\pi \left( {\left( {{e^{ - \frac{{{\rho _b}{\varepsilon _j}}}{{{\rho _s}{s_k}d_0^\alpha }}\sum\limits_{n = 0}^N {{q_n}{c_n}} }} - 1} \right)d_0^2 + \frac{{{\rho _b}{\varepsilon _j}}}{{{\rho _s}{s_k}}}\sum\limits_{n = 0}^N {{q_n}{c_n}} \sum\limits_{l = 1}^L {{\beta _l}{e^{ - \frac{{{t_l}{\rho _b}{\varepsilon _j}}}{{{\rho _s}{s_k}d_0^\alpha }}\sum\limits_{n = 0}^N {{q_n}{c_n}} }}} } \right)}},
\end{align}
where $K$ is a complexity-accuracy tradeoff parameter, ${\omega _K} = \frac{\pi }{K}$, ${\varphi _k} = \cos \left( {\frac{{2k - 1}}{{2K}}\pi } \right)$, ${s_k} = \frac{1}{2}\left( {{\varphi _k} + 1} \right)$, and ${\eta _k} = \frac{{{\omega _K}}}{2}\sqrt {1 - \varphi _k^2} \left( {\frac{{{\rho _p}}}{{{\rho _s}{s_k}}} + \delta } \right){a_\ell }\rho _s^\delta s_k^{\delta  - 1}{e^{ - {a_\ell }\rho _s^\delta s_k^\delta {e^{ - \frac{{{\rho _p}}}{{{\rho _s}{s_k}}}}}}}$.

Substituting \eqref{Q_1_temp} and \eqref{GC Phi} into \eqref{CDF_X_m_wave_GC_1} and applying ${\varepsilon _{\max }} \to {\varepsilon _j}$, based on \eqref{Pout exact 1}, we obtain the closed-form expression of the outage probability at the $m$-th user as follows:
\begin{align} \label{Pout_exact_ordered}
&P_m = {\psi _m}\sum\limits_{p = 0}^{M - m} {
   {M - m}  \choose
   p  } \frac{{{{\left( { - 1} \right)}^p}}}{{m + p}}{\sum _{{q_0} +  \cdots  + {q_N} = m + p}}{
   {m + p}  \choose
   {{q_0} +  \cdots  + {q_N}}  }\left( {\prod\limits_{n = 0}^N {b_n^{{q_n}}} } \right)\nonumber\\
&  \times \left[ {e^{ - {a_\ell }\rho _s^\delta {e^{ - \frac{{{\rho _p}}}{{{\rho _s}}}}} - \frac{{{\varepsilon_{\max}}\sum\limits_{n = 0}^N {{q_n}{c_n}} }}{{{\rho _s}}} - {\lambda _b}\pi \left( {\left( {{e^{ - \frac{{{\rho _b}{\varepsilon_{\max}}}}{{{\rho _s}d_0^\alpha }}\sum\limits_{n = 0}^N {{q_n}{c_n}} }} - 1} \right)d_0^2 + \frac{{{\rho _b}{\varepsilon_{\max}}}}{{{\rho _s}}}\sum\limits_{n = 0}^N {{q_n}{c_n}} \sum\limits_{l = 1}^L {{\beta _l}{e^{ - \frac{{{t_l}{\rho _b}{\varepsilon_{\max}}}}{{{\rho _s}d_0^\alpha }}\sum\limits_{n = 0}^N {{q_n}{c_n}} }}} } \right)}}\right.\nonumber\\
&\left. { + \sum\limits_{k = 1}^K {{\eta _k}} {e^{ - \frac{{{\rho _p} + {\varepsilon_{\max}}\sum\limits_{n = 0}^N {{q_n}{c_n}} }}{{{\rho _s}{s_k}}} - {\lambda _b}\pi \left( {\left( {{e^{ - \frac{{{\rho _b}{\varepsilon_{\max}}}}{{{\rho _s}{s_k}d_0^\alpha }}\sum\limits_{n = 0}^N {{q_n}{c_n}} }} - 1} \right)d_0^2 + \frac{{{\rho _b}{\varepsilon_{\max}}}}{{{\rho _s}{s_k}}}\sum\limits_{n = 0}^N {{q_n}{c_n}} \sum\limits_{l = 1}^L {{\beta _l}{e^{ - \frac{{{t_l}{\rho _b}{\varepsilon_{\max}}}}{{{\rho _s}{s_k}d_0^\alpha }}\sum\limits_{n = 0}^N {{q_n}{c_n}} }}} } \right)}}} \right].
\end{align}

\section{Diversity Analysis}
Based on the analytical results for the outage probability in \eqref{Pout_exact_ordered}, we aim to provide asymptotic diversity analysis for the ordered NOMA users. The diversity order of the user's outage probability is defined as
\begin{align}\label{Diversity 1}
d =  - \mathop {\lim }\limits_{\rho_s  \to \infty } \frac{{\log {P_m}\left( \rho_s  \right)}}{{\log \rho_s }}.
\end{align}
\subsection{Fixed Transmit Power at Primary  Transmitters}
In this case, we examine the diversity with the fixed transmit SNR at the PTs ($\rho_b$), while the transmit SNR of secondary BS ($\rho_s$) and the maximum permissible interference constraint at the PRs ($\rho_p$) go to the infinity. Particularly, we assume $\rho_p$ is proportional to $\rho_s$, i.e.  ${\rho _p} = \kappa {\rho _s}$, where $\kappa$ is a positive scaling factor. This assumption applies to the scenario where the PRs can tolerate a large amount of interference from the secondary BS and the target data rate is relatively small in the PN. Denote ${\gamma _{{t^*}}} = \frac{{{\gamma _t}}}{{{\rho _s}}}=min \left\{ {\frac{\kappa }{{\mathop {\max }\limits_{\ell  \in {\Phi _\ell }} {{\left| {{g_\ell }} \right|}^2}}},1} \right\}$, similar to \eqref{order stacastics}, the ordered CDF has the relationship with unordered CDF as
\begin{align}\label{Diversity case 1_1}
F_{\left. {{X_m}} \right|{I_B},{\gamma _{{t^*}}}}^\infty \left( {{y^*}} \right) = {\psi _m}\sum\limits_{p = 0}^{M - m} {
 M - m \choose
 p } \frac{{{{\left( { - 1} \right)}^p}}}{{m + p}}{\left( {F_{{{\left| {{{\tilde h}}} \right|}^2}}^\infty \left( {{y^*}} \right)} \right)^{m + p}},
\end{align}
where $y^* = \frac{{\left( {{\rho _b}{I_B} + 1} \right){\varepsilon _j}}}{{{\rho _s}{\gamma _{{t^*}}}}}$.
When ${\rho _s} \to \infty$, we observe that $y^* \to 0$. In order to investigate an insightful expression to obtain the diversity order, we use Gaussian-Chebyshev quadrature and $1 - {e^{ - y^*}} \approx y^*$ to approximate \eqref{CDF_X_m_wave} as
\begin{align}\label{CDF_X_m_wave appro}
F_{{{\left| {{{\tilde h}}} \right|}^2}}^\infty \left( {{y^*}} \right) \approx \sum\limits_{n = 1}^N {{\chi _n}{y^*}} ,
\end{align}
where ${\chi _n} = {\omega _N}\sqrt {1 - \phi _n^2} \left( {{\phi _n} + 1} \right){c_n}$. Substituting \eqref{CDF_X_m_wave appro} into \eqref{Diversity case 1_1}, since $y^* \to 0$, we obtain
\begin{align}\label{Diversity fixed 1}
F_{_{\left. {{X_m}} \right|{I_B},{\gamma _{{t^*}}}}}^\infty \left( \varepsilon _j \right) = \xi {\left( \frac{{\left( {{\rho _b}{I_B} + 1} \right){\varepsilon _j}}}{{{\rho _s}{\gamma _{{t^*}}}}}\right)^m} + o\left( {{{\left( \frac{{\left( {{\rho _b}{I_B} + 1} \right){\varepsilon _j}}}{{{\rho _s}{\gamma _{{t^*}}}}} \right)}^m}} \right),
\end{align}
where $\xi  = \frac{{{\psi _m}{{\left( {\sum\limits_{n = 1}^N {{\chi _n}} } \right)}^m}}}{m}$. Based on \eqref{Pout exact 1}, \eqref{CDF_X_m_2}, and \eqref{Diversity fixed 1}, the asymptotic outage probability is given by
\begin{align}\label{CDF asym fixed 2}
P_{{m_F}}^\infty  \approx \frac{1}{{\rho _s^m}}\underbrace {\int_0^\infty  {\int_0^\infty  {\xi {{\left( {\frac{{\left( {{\rho _b}x + 1} \right){\varepsilon _{\max }}}}{z}} \right)}^m}{f_{{I_B}}}\left( x \right){f_{{\gamma _{{t^*}}}}}\left( z \right)dxdz} } }_C,
\end{align}
where ${{f_{{\gamma _{{t^*}}}}}}$ the PDF of ${\gamma _{t^*}}$. Since $C$ is a constant independent of ${{\rho _s}}$, \eqref{CDF asym fixed 2} can be expressed as follows:
\begin{align}\label{CDF asym fixed 3}
P_{{m_F}}^\infty  = \frac{1}{{\rho _s^m}}C + o\left( {{\rho _s}^{ - m}} \right),
\end{align}

Substituting \eqref{CDF asym fixed 3} into \eqref{Diversity 1}, we obtain the diversity order of this case is $m$. This can be explained as follows. Note that SIC is applied at the ordered SUs. For the first user with the poorest channel gain, no interference cancelation is operated at the receiver, therefore its diversity gain is one. While for the $m$-th user, since the interferences from all the other $(m-1)$ users are canceled, it obtains a diversity of $m$.
\vspace{-0.5cm}
\subsection{ Transmit Power of Primary Transmitters Proportional to that of Secondary Ones}
In this case, we examine the diversity with the transmit SNR at the PTs ($\rho_b$) is proportional to the transmit SNR of secondary BS ($\rho_s$).  Particularly, we assume ${\rho _b} = \nu {\rho _s}$, where $\nu$ is a positive scaling factor. We still assume $\rho_p$ is proportional to $\rho_s$. Applying ${\rho _s} \to \infty $, ${\rho _p} = \kappa {\rho _s}$ and ${\rho _b} = \nu {\rho _s}$ to \eqref{Pout_exact_ordered}, we obtain the asymptotic outage probability of the $m$-th user in this case as follows:
\begin{align} \label{Pout_exact_ordered asymptotic}
P_{{m_P}}^\infty \approx & {\psi _m}\sum\limits_{p = 0}^{M - m} {
   {M - m}  \choose
   p  } \frac{{{{\left( { - 1} \right)}^p}}}{{m + p}}{\sum _{{q_0} +  \cdots  + {q_N} = m + p}}{
   {m + p}  \choose
   {{q_0} +  \cdots  + {q_N}}  }\left( {\prod\limits_{n = 0}^N {b_n^{{q_n}}} } \right)\nonumber\\
&  \times \left[ {{e^{ - a_\ell ^\infty {e^{ - \kappa }} - {\lambda _b}\pi \left( {\left( {{e^{ - \frac{{\nu {\varepsilon _{\max }}}}{{d_0^\alpha }}\sum\limits_{n = 0}^N {{q_n}{c_n}} }} - 1} \right)d_0^2 + \nu {\varepsilon _{\max }}\sum\limits_{n = 0}^N {{q_n}{c_n}} \sum\limits_{l = 1}^L {{\beta _l}{e^{ - \frac{{{t_l}\nu {\varepsilon _{\max}}}}{{d_0^\alpha }}\sum\limits_{n = 0}^N {{q_n}{c_n}} }}} } \right)}}} \right.\nonumber\\
&\left. { + \sum\limits_{k = 1}^K {\eta _k^\infty } {e^{ - \frac{\kappa }{{{s_k}}} - {\lambda _b}\pi \left( {\left( {{e^{ - \frac{{\nu {\varepsilon _{\max }}}}{{{s_k}d_0^\alpha }}\sum\limits_{n = 0}^N {{q_n}{c_n}} }} - 1} \right)d_0^2 + \frac{{\nu {\varepsilon _{\max }}}}{{{s_k}}}\sum\limits_{n = 0}^N {{q_n}{c_n}} \sum\limits_{l = 1}^L {{\beta _l}{e^{ - \frac{{{t_l}\nu {\varepsilon _{\max}}}}{{{s_k}d_0^\alpha }}\sum\limits_{n = 0}^N {{q_n}{c_n}} }}} } \right)}}} \right].
\end{align}
where $a_\ell ^\infty  = \frac{{\delta \pi {\lambda _\ell }\Gamma \left( \delta  \right)}}{{{\kappa ^\delta }}}$ and $\eta _k^\infty  = \frac{{{\omega _K}}}{2}\sqrt {1 - \varphi _k^2} \left( {\frac{\kappa }{{{s_k}}} + \delta } \right)a_\ell ^\infty s_k^{\delta  - 1}{e^{ - a_\ell ^\infty s_k^\delta {e^{ - \frac{\kappa }{{{s_k}}}}}}}$.

It is observed that $P_{{m_P}}^\infty$ is a constant   independent of $\rho_s$. Substituting \eqref{Pout_exact_ordered asymptotic} into \eqref{Diversity 1}, we find that asymptotically there is an error floor for the outage probability of SUs.

\vspace{-0.8cm}
\section{Numerical Results}
In this section, numerical results are presented to verify the accuracy of the analysis as well as to obtain more important insights for NOMA in large-scale CR networks. In the considered network, the radius of the guard zone is assumed to be $d_0=2$ m. The Gaussian-Chebyshev parameters are chosen with $N=5$, $K=10$, and $L=10$. Monte Carlo simulation results are marked as ``$\bullet$" to verify our derivation.
\begin{figure}[t!]
\centering
\subfigure[For different user zone, with $\lambda_b=10^{-3}$, $\lambda_\ell=10^{-3}$, $\kappa=1$, $\alpha=4$, $\rho_b=20$ dB, and $M=3$.]{
\label{P_m_out_fixed 1}
\includegraphics[width=3.3 in]{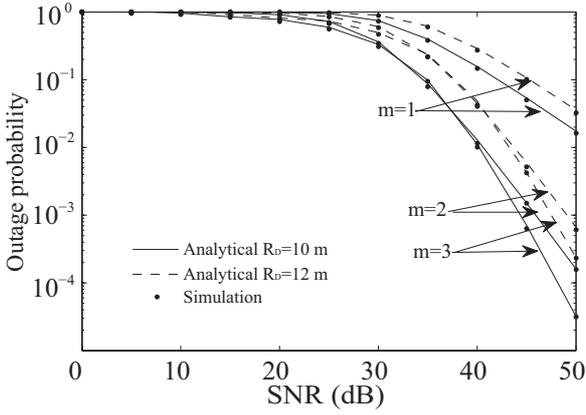}}
\;\;\;\;\;\;\subfigure[For different $\alpha$, with $\lambda_b=10^{-3}$, $\lambda_\ell=10^{-3}$, $\kappa=1$, $R_D=5$~m, $\rho_b=20$ dB, and $M=2$.]{
\label{P_m_out_fixed 2}
\includegraphics[width=3.3 in]{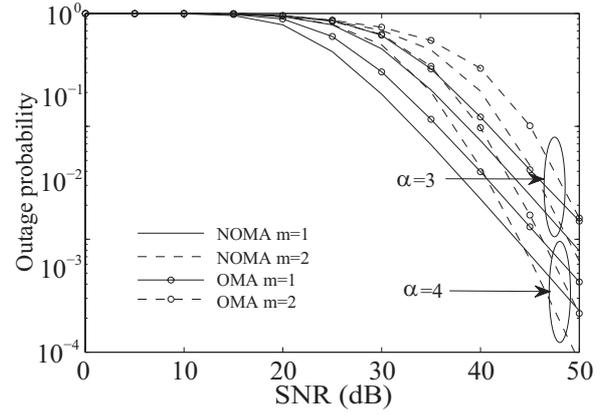}}
\caption{Outage probability of the $m$-th user versus $\rho_s$ of the first scenario.}
\label{P_m_out_fixed}
\end{figure}

\begin{figure}[t!]
\centering
\subfigure[For different density of PTs and PRs, with $\alpha=4$, $\kappa=1$, $\nu=1$, $R_D=10$ m, $\rho_b=\nu\rho_s$, and $M=2$.]{
\label{P_m_out_propotional 1}
\includegraphics[width=3.3 in]{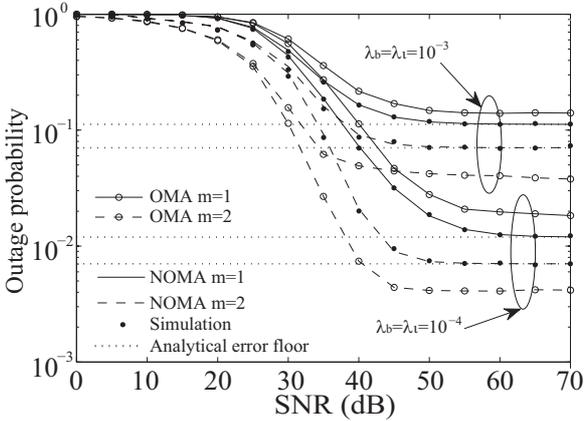}}
\;\;\;\;\;\;\subfigure[For different $\nu$, with $\alpha=4$, $\lambda_b=10^{-4}$, $\lambda_\ell=10^{-4}$, $\kappa=0.5$, $R_D=10$ m, $\rho_b=\nu\rho_s$, and $M=2$.]{
\label{P_m_out_propotional 2}
\includegraphics[width=3.3 in, height=2.2in]{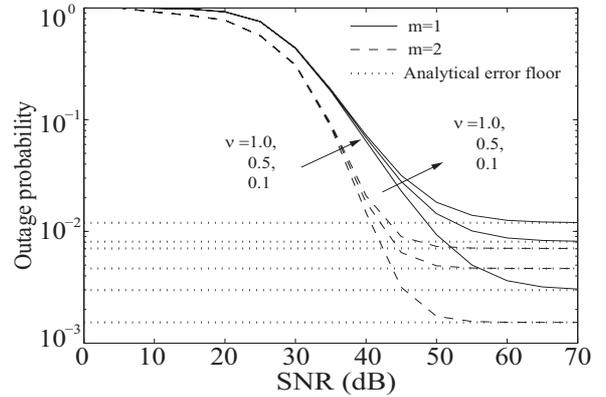}}
\caption{Outage probability of the $m$-th user versus $\rho_s$ of the second scenario.}
\label{P_m_out_propotional}
\end{figure}
Fig. \ref{P_m_out_fixed} plots the outage probability of the $m$-th user for the first scenario when $\rho_b$ is fixed and $\rho_p$ is proportional to $\rho_s$. In Fig. \ref{P_m_out_fixed 1}, the power allocation coefficients are $a_1=0.5$, $a_2=0.4$ and $a_3=0.1$. The target data rate for each user is assumed to be all the same as $R_1=R_2=R_3=0.1$ bit per channel use (BPCU). The dashed and solid curves are obtained from the analytical results derived in \eqref{Pout_exact_ordered}.  Several observations can be drawn as follows:  1) Reducing the coverage of the secondary users zone $\mathcal{D}$ can achieve a lower outage probability because of a smaller path loss. 2) The ordered users with different channel conditions have different decreasing slope because of different diversity orders, which verifies the derivation of \eqref{CDF asym fixed 2}.
In Fig. \ref{P_m_out_fixed 2}, the power allocation coefficients are $a_1=0.8$ and $a_2=0.2$. The target rate is $R_1=1$ and $R_2=3$ BPCU. The performance of a conventional OMA is also shown in the figure as a benchmark for comparison. It can be observed that for different values of the path loss, NOMA can achieve a lower outage probability than the conventional OMA.

Fig. \ref{P_m_out_propotional} plots the outage probability of the $m$-th user for the second scenario when both $\rho_b$ and $\rho_p$ are proportional to $\rho_s$. The power allocation coefficients are $a_1=0.8$ and $a_2=0.2$. The target rates are $R_1=R_2=0.1$ BPCU. The dashed and solid curves are obtained from the analytical results derived in \eqref{Pout_exact_ordered}. One observation is that error floors exist in both Figs. \ref{P_m_out_propotional 1} and \ref{P_m_out_propotional 2}, which verifies the asymptotic results in \eqref{Pout_exact_ordered asymptotic}. Another observation is that user two ($m=2$) outperforms user one ($m=1$). The reason is that for user two, by applying SIC, the interference from user one is canceled. While for user one, the interference from user two still exists. In Fig. \ref{P_m_out_propotional 1}, it is shown that the error floor become smaller when $\lambda_b$ and $\lambda_\ell$ decrease, which is due to less interference from PTs and the relaxed interference power constraint at the PRs. It is also worth noting that with these system parameters, NOMA outperforms OMA for user one while OMA outperforms NOMA for user two, which indicates the importance of selecting appropriate power allocation coefficients and target data rates for NOMA. In Fig. \ref{P_m_out_propotional 2}, it is observed that the error floors become smaller as $\nu$ decreases. This is due to the fact that smaller $\nu$ means a lower transmit power of PTs, which in turn reduces the interference at SUs.

\vspace{-0.8cm}
\section{Conclusions}
In this paper, we have studied   non-orthogonal multiple access (NOMA) in large-scale underlay cognitive radio networks with randomly deployed users. Stochastic geometry tools were used to evaluate the outage performance of the considered network. New closed-form expressions were derived for the outage probability. Diversity order of NOMA users has been analyzed in two situations based on the derived outage probability. An important future direction is to optimize the power allocation coefficients to further improve the performance gap between NOMA and conventional MA in CR networks.

\vspace{-0.5cm}
\numberwithin{equation}{section}
\section*{Appendix~A: Proof of Theorem~\ref{theorem:A}} \label{Appendix:A}
\renewcommand{\theequation}{A.\arabic{equation}}
\setcounter{equation}{0}

The CDF of ${\gamma _t}$ is given by
\begin{align}\label{CDF_Pt}
{F_{{{\gamma _t}}}}\left( x \right) &= \Pr \left\{ {\min \left\{ {\frac{{{\rho _p}}}{{\mathop {\max }\limits_{\ell  \in {\Phi _\ell }} {{\left| {{g_\ell }} \right|}^2}}},{\rho _s}} \right\} \le x} \right\}\nonumber\\
&= \Pr \left\{ {\mathop {\max }\limits_{\ell  \in {\Phi _\ell }} {{\left| {{g_\ell }} \right|}^2} \ge \max \left\{ {\frac{{{\rho _p}}}{x},\frac{{{\rho _p}}}{{{\rho _s}}}} \right\}} \right\}+ \Pr \left\{ {\mathop {\max }\limits_{\ell  \in {\Phi _\ell }} {{\left| {{g_\ell }} \right|}^2} \le \frac{{{\rho _p}}}{{{\rho _s}}},{\rho _s} \le x} \right\}\nonumber\\
&= 1 - U\left( {{\rho _s} - x} \right)\underbrace {\Pr \left\{ {\mathop {\max }\limits_{\ell  \in {\Phi _\ell }} {{\left| {{g_\ell }} \right|}^2} \ge \frac{{{\rho _p}}}{x}} \right\}}_\Omega.
\end{align}

Denote $\bar \Omega  = 1 - \Omega $, we express $\bar \Omega$ as follows:
\begin{align}\label{CDF_Q2_1}
\bar \Omega &= \Pr \left\{ {\mathop {\max }\limits_{\ell  \in {\Phi _\ell }} {{\left| {{g_\ell }} \right|}^2} \le {\frac{{{\rho _p}}}{x}} } \right\} = {E_{{\Phi _\ell }}}\left\{ {{\prod _{\ell  \in {\Phi _\ell }}}\Pr \left\{ {{{\left| {{{\hat g}_\ell }} \right|}^2} \le \frac{{\left( {1 + d_\ell ^\alpha } \right){\rho _p}}}{x}} \right\}} \right\}\nonumber\\
&= {E_{{\Phi _\ell }}}\left\{ {{\prod _{\ell  \in {\Phi _\ell }}}{F_{{{\left| {{{\hat g}_\ell }} \right|}^2}}}\left( {\frac{{\left( {1 + d_\ell ^\alpha } \right){\rho _p}}}{x}} \right)} \right\}.
\end{align}

Applying the generating function, we rewrite \eqref{CDF_Q2_1} as follows:
\begin{align}\label{CDF_Q2_2}
\bar \Omega  =& {\rm{ }}\exp \left[ { - {\lambda _\ell }\int\limits_{{R^2}} {\left( {1 - {F_{{{\left| {{{\hat g}_\ell }} \right|}^2}}}\left( {\left( {1 + d_\ell ^\alpha } \right)\mu } \right)} \right)rdr} } \right]=\exp \left[ { - 2\pi {\lambda _\ell }{e^{ - \mu }}\int_0^\infty  {r{e^{ - \mu {r^\alpha }}}dr} } \right].
\end{align}
Applying \cite[ Eq. (3.326.2)]{gradshteyn},  we obtain
\begin{align}\label{CDF_Q2_3}
\Omega  = 1 - \bar \Omega  = 1 - {e^{ - \frac{{{e^{ - \mu }}\delta \pi {\lambda _\ell }\Gamma \left( \delta  \right)}}{{{\mu ^\delta }}}}},
\end{align}
where $\Gamma(\cdot)$ is Gamma function. Substituting \eqref{CDF_Q2_3} into \eqref{CDF_Pt}, and taking the derivative, we obtain the PDF of ${\gamma _t}$ in \eqref{PDF Pt}. The proof is completed.

%
\vspace{-0.5cm}
\bibliographystyle{IEEEtran}
\bibliography{mybib}

\end{document}